\documentclass{elsart}
\usepackage{natbib}

\newcommand{\ltsimeq}{\raisebox{-0.6ex}{$\,\stackrel 
        {\raisebox{-.2ex}{$\textstyle <$}}{\sim}\,$}}

\begin{document}
\runauthor{Brand}
\begin{frontmatter}
\title{Using radio galaxies to find super-structures}

\author{Kate Brand$^{1}$, Steve Rawlings$^{1}$, Gary J. Hill$^{2}$, Mark Lacy$^{3}$ }\\
\small{$^1$ \it{Astrophysics, Department of Physics, Keble Road, Oxford OX1 3RH, UK} \\
$^2$ \it{University of Texas at Austin, Austin, Texas 78712, USA}\\
$^3$ \it{SIRTF Science Center; MS-220-6; California Institute of Technology, 1200 E. California Boulevard, Pasadena, CA 91125}\\}


\begin{abstract}

Radio galaxies are excellent at tracing large-scale structure due to their high bias. We present new results from the TONS08 radio galaxy redshift survey. We find unequivocal evidence for a huge (at least 80 $\times$ 80 $\times$ 100 $\rm {Mpc}^3$) super-structure at redshift $z=0.27$, confirming tentative evidence for such a structure from the 7C redshift survey (7CRS). A second, newly discovered super-structure is also tentatively found at redshift 0.35 (of dimensions at least 100 $\times$ 100 $\times$ 100 $\rm{Mpc}^3$). Out of the total sample size of 84 radio galaxies, at least 25 are associated with the two super-structures. We use quasi-linear structure formation theory to estimate the number of such structures expected in the TONS08 volume if the canonical value for radio galaxy bias is assumed. Under this assumption, the structures represent $\approx$ 4-5$\sigma$ peaks in the primordial density field and their expected number is low ($\sim 10^{-2}-10^{-4}$). 
Fortunately, there are several plausible explanations (many of which are testable) for these low probabilities in the form of potential mechanisms for boosting the bias on large scales. These include: the association of radio galaxies with highly biased rich clusters in super-structures, enhanced triggering by group/group mergers, and enhanced triggering and/or redshift space distortion in collapsing systems as the growth of super-structures moves into the non-linear regime. Similar structures could have been missed in previous surveys because of the effects of Poisson-sampling fluctuations.
\end{abstract}
\begin{keyword}
radio continuum:$\>$galaxies -- galaxies:$\>$active -- cosmology:$\>$observations -- cosmology:$\>$large-scale structure of the Universe
\end{keyword}

\end{frontmatter}

\section{Introduction}

Super-structures are the largest known structures in the Universe. They have evolved from the rare ($>$3$\sigma$) peaks in the initial density field at recombination \citep{hna}. Cosmological biasing predicts that large-scale dark-matter fluctuations help push small-scale fluctuations over the density contrast required for gravitational collapse to occur \citep{kai}, meaning that aggregates of rich clusters are expected to trace super-structures in the same way that aggregates of galaxies trace rich clusters. 

Radio galaxies are ideal tracers of super-structures because they are associated with massive elliptical galaxies, which themselves tend to reside in clusters. Because they are biased tracers of the mass \citep{pd}, they therefore allow us to search large sky areas efficiently. It is only with the advent of radio surveys such as NVSS \citep{con} and FIRST \citep{bwh} that we can compile the large area samples of faint radio galaxies that we need to study the large-scale structure. 
\section{The TONS08 survey}

The TONS08 (Texas-Oxford NVSS Structure 08$^h$ region) survey is in one of the areas covered by the 7CRS \citep{wil02} and the Texox-1000 (TOOT) survey \citep{hr}. Unlike 7CRS or TOOT, the TONS08 survey is selected at 1.4 GHz from the NVSS survey. It therefore goes to fainter radio flux densities than TOOT. It also has an optical magnitude limit imposed on it (see \citep{brand}).  
TONS08 was specifically designed to seek further evidence for a giant super-structure found in this region of sky by analysis of the redshift distribution of in the 7C-II survey \citep{rhw}. The redshift distribution in the 7C-II survey revealed a prominent spike at $z$ $\approx$ 0.25 of five radio galaxies from the 7CRS. By going to fainter radio flux densities ($S_{\rm{1.4}}\ge$ 3 mJy) and optical limits ($E\approx R\approx$19.5), the TONS08 sample was optimised for looking at clustering of objects in a region of moderate (0 $\ltsimeq$ $z$ $\ltsimeq$ 0.5) redshifts.

The selection criteria for the TONS08 survey was based on cross-matching positions of objects in radio (NVSS) and optical (APM) surveys. To ensure a complete, flux-density-limited survey, we plotted radio contours from the FIRST survey over optical POSS-II images, identifying radio galaxy identifications by eye. 

\begin{figure}
\begin{center}
\setlength{\unitlength}{1mm}
\begin{picture}(80,40)
\put(-10,65){\includegraphics{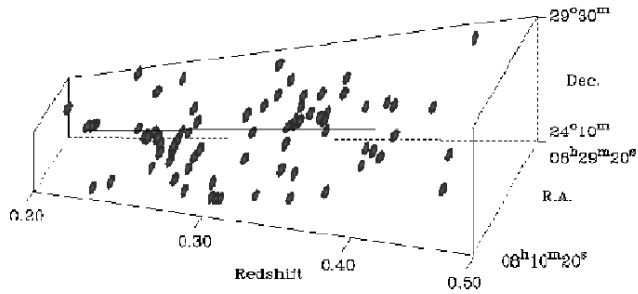}}
\end{picture}
{\caption[3-D distribution of TONS08 sample]{\label{fig:3d} Representation of the three-dimensional distribution of the TONS08 sample between $z$=0.2 and $z$=0.5. Note the effects of the selection function in redshift plotted in Fig.~\ref{fig:zdistn_rlf}. }} 
\end{center}
\end{figure}

We obtained optical spectra for all the 84 radio galaxies in TONS08. All but three had secure redshifts (see \citep{brand} for spectra and more details). A representation of the three-dimensional distribution of the TONS08 sample is shown in Fig.~\ref{fig:3d}. Clustering of objects can be clearly seen at a redshift of $z$ $\approx$ 0.27 as well as a second clustering at $z \approx$ 0.35.

\section{A model redshift distribution}

\begin{figure}
\begin{center}
\setlength{\unitlength}{1mm}
\begin{picture}(150,65)
\put(30,-10){\includegraphics{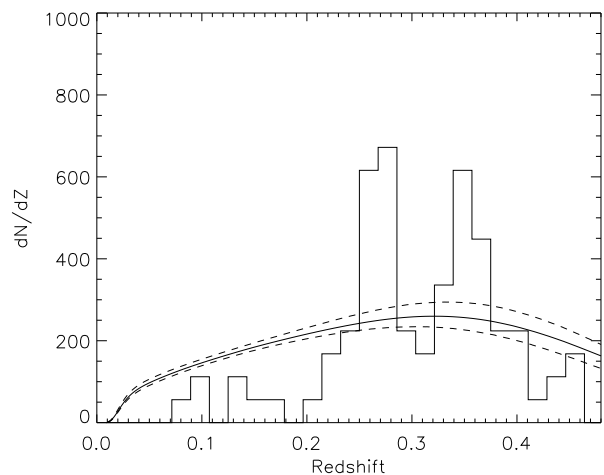}}
\end{picture}
\end{center}
{\caption[junk]{\label{fig:zdistn_rlf} The redshift distribution of the TONS08 sample with the model redshift distribution overplotted (solid line). The $\pm 1\sigma$ errors on the model are overplotted (dashed lines).
}}
\end{figure}

In order to determine how significant the redshift spikes are, and to determine the radio galaxy overdensity, we constructed a model redshift distribution. This was obtained by fitting a model bivariate (radio and optical) luminosity function (BLF) to data from radio galaxies in the 2dFGRS \citep{sad} and integrating it over the optical and radio selection limits. A much more accurate and representative redshift distribution is obtained because the sky area is $\approx$ 13-times larger than the TONS08 area. The model redshift distribution, normalised to the TONS08 sky area is shown in Fig.~\ref{fig:zdistn_rlf}. Also plotted is the binned TONS08 redshift distribution. The $\pm 1 \sigma$ errors on the model redshift distribution were found by performing a Monte-Carlo simulation on the parameters of the model. 

The overdensities in Fig.~\ref{fig:3d} can be clearly seen as redshift spikes in Fig.~\ref{fig:zdistn_rlf}. To determine whether they are significant, we used a simple method using Poisson statistics in each redshift bin to work out the probability that the number of galaxies could be greater than the actual number given the number predicted by the model. We excluded the four 7C-II objects at redshift $z\approx$ 0.27 to find independent evidence for redshift spikes, and found that the redshift spike at $z\approx$ 0.27 would be expected in only 0.3 per cent of random realisations. A redshift spike as large as the one at $z\approx$ 0.35 would , as a new feature in any one of the bins, be expected in 4 per cent of random realisations. We performed the same calculations using redshift distributions corresponding to the $+ 1 \sigma$ error on the model redshift distribution. This increases the chance of there being such a peak to 0.6 per cent and 14 per cent of random realisations for the $z\approx$ 0.27 and $z\approx$ 0.35 redshift spikes respectively. We conclude that the redshift spike at $z$=0.27 is significant, but that evidence for the $z$=0.35 redshift spike is less secure.

To match the data to theoretical overdensities, we chose to calculate overdensities in 50-Mpc radius spheres centred at the peaks of the redshift spikes. In this case we found 13 radio galaxies in the $z$=0.27 super-structure and 12 in the $z$=0.35 super-structure. We find an overdensity of $\delta_{\rm{gal}}\approx 2.27\pm^{0.59}_{0.29}$ for the $z$=0.27 redshift peak and $\delta_{\rm{gal}}\approx 3.41\pm^{0.40}_{0.31}$ for the z=0.35 redshift peak. The errors quoted are 68 per cent confidence limits. The overdensity for the $z$=0.35 super-structure is larger than that of the $z$=0.27 super-structure even though the model redshift distribution is higher at $z$=0.35. This is because the 50-Mpc sphere takes up less of the survey sky area for the $z$=0.35 super-structure.

\section{How probable is it that we find them?}

\begin{figure*}
\begin{center}
\setlength{\unitlength}{1mm}
\begin{picture}(150,70)
\put(10,55){{\textbf{$z$=0.27 Super-Structure}}}
\put(0,-5){\includegraphics{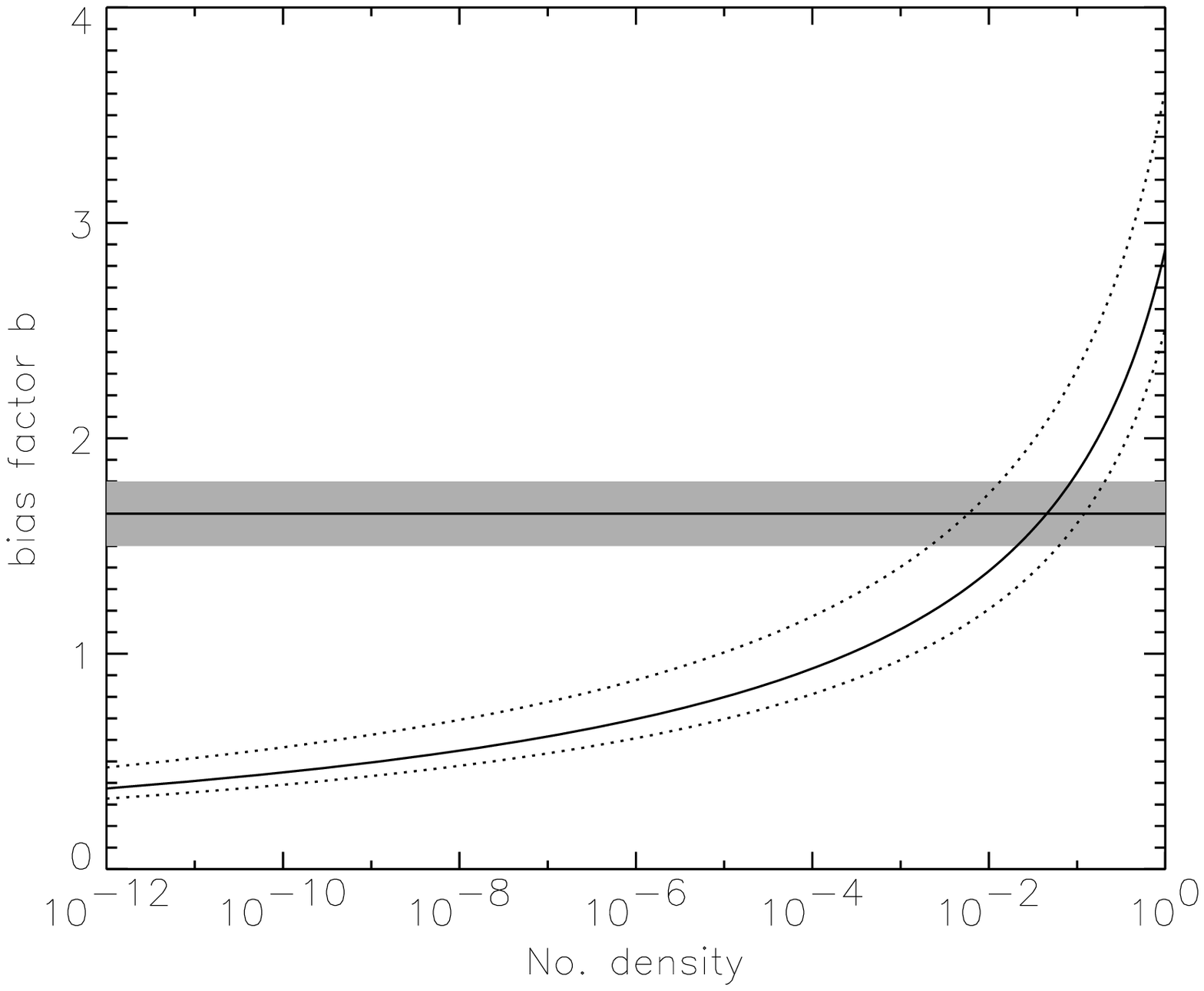}}
\put(80,55){{\textbf{$z$=0.35 Super-Structure}}}
\put(70,-5){\includegraphics{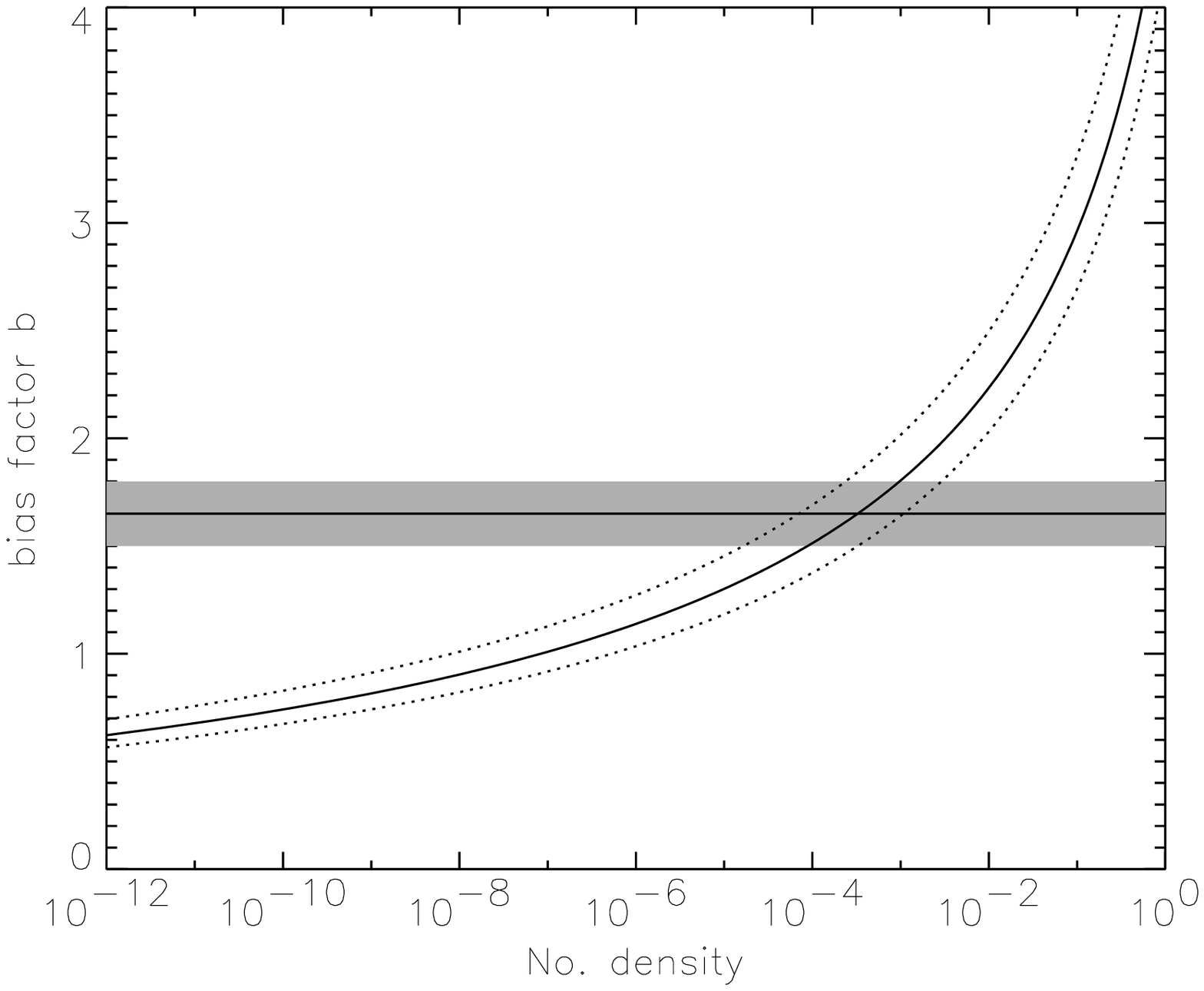}}

\end{picture}
\end{center}
{\caption[junk]{\label{fig:prob_bias} The probability of intercepting a super-structure of radius 50 Mpc and the galaxy overdensity of the $z$=0.27 super-structure (left) and $z$=0.35 super-structure (right) against the radio galaxy bias factor; $\sigma_{8h^{-1}}$ is assumed to be 0.94. Overplotted are the 68 per cent confidence limits of the galaxy overdensities which gives us an indication of the effects of Poisson-sampling fluctuations. A local value for the radio galaxy bias b=1.65$\pm$0.15,and its uncertainty, is plotted. 
}}
\end{figure*}

To determine how probable it is that we detected these two super-structures, we calculated the probability of the TONS08 survey intercepting such structures given the cosmological parameters and different degrees of radio galaxy bias. This was done using standard structure formation theory (see \citep{brand,bar} for more details). We used a log-normal (LN) random field \citep{cj} to make some attempt at correcting the mass overdensity to a non-linear-corrected mass overdensity. Fig.~\ref{fig:prob_bias} shows the probability of intercepting a super-structure of radius 50-Mpc and the galaxy overdensity of the two super-structures. 
We assume the local value for the radio galaxy bias factor is $b$=1.65$\pm0.15$. This is determined by multiplying the relative bias factor of radio galaxies to optical galaxies (1.9/1.3=1.5$\pm$ 6 per cent calculated by \citep{pd}) with the bias factor of optical galaxies relative to the dark matter (1.1$\pm$7 per cent calculated by \citep{lah}, and adding the errors in quadrature.

\section{Discussion and conclusions}

If we assume a relatively high value for $\sigma_{8h^{-1}}$ (0.94) and that the radio galaxy bias parameter is the same as for smaller scales in the local Universe, Fig.~\ref{fig:prob_bias} forces us to conclude that we have been very fortunate in our choice of survey area and have found two rare (4 and 5 $\sigma$) fluctuations of which we would expect only 0.05 and 3$\times 10^{-4}$ in our sample volume (for the $z$=0.27 and $z=0.35$ super-structures respectively). 

To explain these results in the framework of standard $\Lambda$CDM model and the inflationary (i.e.\ Gaussian fluctuation) paradigm for structure formation, we need to determine why there is such a discrepancy between this theoretical result and the fact that we have found two super-structures in such a small survey volume. 

The probability of seeing two such fluctuations in TONS08 is strongly dependent on how representative the survey region is of the Universe. If, by targeting a structure discovered in the 7CRS, we have focused on a particularly unusual part of the Universe, then the probability of seeing two super-structures is roughly equal to the probability of finding only the $z$=0.35 super-structure, namely $\sim 10^{-4}$. If, however, the TONS08 target region turns out to be a reasonably representative one, then given our assumptions, the probability of seeing two super-structures is $\sim 10^{-5}$. 

The most likely explanation of our results is that radio galaxies in super-structures are extremely biased tracers of mass (i.e. more biased than local radio galaxies). This could be due to redshift evolution of the bias factor and/or due to different mechanisms being at work on super-structure scales. Assuming an evolution of the bias factor with redshift that is consistent with previous studies \citep{bw,cro}, some boosting of bias in super-structures is still needed. We suggest that the most promising way of reconciling the low probability of TONS08 intercepting both super-structures with the fact that we have detected them is either that the radio galaxies within the super-structure have a higher bias factor than is derived for large samples of radio galaxies at $z\sim$0 \citep{pn} and $z\sim$1 \citep{bw} and/or they trace collapsing systems. There are several plausible explanations for this. Perhaps collapsed sub-structure within the super-structure, i.e. rich clusters, hosts the radio galaxies and the high bias simply reflects that of the rare rich clusters. Perhaps there is an enhanced group-group merger rate within the super-structure which is enhancing radio galaxy triggering. Perhaps as the super-structure goes non-linear, redshift space distortions and/or enhanced triggering due to high velocities boost the bias. The good news is that it may be possible to discriminate between these possibilities by studying the environment of the radio galaxies within the two super-structures and we are in the process of doing this.

\end{document}